# IDTxl: The Information Dynamics Toolkit xl: a Python package for the efficient analysis of multivariate information dynamics in networks

Patricia Wollstadt[1], Joseph T. Lizier[2], Raul Vicente[3], Conor Finn[2, 4], Mario Martinez-Zarzuela[5], Pedro Mediano[6], Leonardo Novelli[2], and Michael Wibral[1, 7, 8]

**1** MEG Unit, Brain Imaging Center, Goethe-University Frankfurt, Fankfurt am Main, Germany **2** Centre for Complex Systems, Faculty of Engineering and IT, The University of Sydney, Sydney, Australia **3** Computational Neuroscience Lab, Institute of Computer Science, Tartu, Estonia **4** Data61, CSIRO, Epping, Australia **5** Communications and Signal Theory and Telematics Engineering, University of Valladolid, Valladolid, Spain **6** Computational Neurodynamics Group, Department of Computing, Imperial College London, London, United Kingdom **7** Max Planck Institute for Dynamics and Self-Organization, Göttingen, Germany **8** Campus Institute for Dynamics of Biological Networks, Georg-August Universität, Göttingen, Germany

## Summary

We present IDTxl (the **I**nformation **D**ynamics **T**oolkit **xl**), a new open source Python toolbox for effective network inference from multivariate time series using information theory, available from GitHub (https://github.com/pwollstadt/IDTxl).

Information theory (Cover & Thomas, 2006; MacKay, 2003; Shannon, 1948) is the mathematical theory of information and its transmission over communication channels. Information theory provides quantitative measures of the information content of a single random variable (entropy) and of the information shared between two variables (mutual information). The defined measures build on probability theory and solely depend on the probability distributions of the variables involved. As a consequence, the dependence between two variables can be quantified as the information shared between them, without the need to explicitly model a specific type of dependence. Hence, mutual information is a model-free measure of dependence, which makes it a popular choice for the analysis of systems other than communication channels.

Transfer entropy (TE) (Schreiber, 2000) is an extension of mutual information that measures the directed information transfer between time series of a source and a target variable. TE has become popular in many scientific disciplines to infer dependencies and whole networks from data. Notable application domains include neuroscience (Wibral, Vicente, & Lindner, 2014) and dynamical systems analysis (Lizier, Prokopenko, & Zomaya, 2014) (see Bossomaier, Barnett, Harré, & Lizier (2016) for an introduction to TE and a comprehensive discussion of its application). In the majority of the applications, TE is used in a bivariate fashion, where information transfer is quantified between all sourcetarget pairs. In a multivariate setting, however, such a bivariate analysis may infer spurious or redundant interactions, where multiple sources provide the same information about the target. Conversely, bivariate analysis may also miss synergistic interactions between multiple relevant sources and the target, where these multiple sources jointly transfer more information into the target than what could be detected from examining source contributions individually. Hence, tools for multivariate TE estimation, accounting for all relevant sources of a target, are required. An exhaustive multivariate approach is computationally intractable, even for a small number of potential sources in the data. Thus, a suitable approximate approach is needed. Although such approaches have been proposed (e.g., Lizier & Rubinov (2012) and Faes, Nollo, & Porta (2011)) and first software implementations exist (Montalto, Faes, & Marinazzo, 2014), there is no current implementation that deals with the practical problems that arise in multivariate TE estimation. These problems include the control of statistical errors that arise from testing multiple potential sources in a data set, and the optimization of parameters necessary for the estimation of multivariate TE.

IDTxl provides such an implementation, controlling for false positives during the selection of relevant sources and providing methods for automatic parameter selection. To estimate multivariate TE, IDTxl utilises a greedy or iterative approach that builds sets of parent sources for each target node in the network through maximisation of a conditional mutual information criterion (Faes et al., 2011; Lizier & Rubinov, 2012). This iterative conditioning is designed to both removes redundancies and capture synergistic interactions in building each parent set. The conditioning thus automatically constructs a non-uniform, multivariate embedding of potential sources (Faes et al., 2011) and optimizes source-target delays (Wibral et al., 2013). Rigorous statistical controls (based on comparison to null distributions from time-series surrogates) are used to gate parent selection and to provide automatic stopping conditions for the inference, requiring only a minimum of user-specified settings.

Following this greedy approach, IDTxl implements further algorithms for network inference (multivariate mutual information, bivariate mutual information, and bivariate transfer entropy), and provides measures to study the dynamics of various information flows on the inferred networks. These measures include active information storage (AIS) (Lizier, Prokopenko, & Zomaya, 2012) for the analysis of information storage within network nodes, and partial information decomposition (PID) (Bertschinger, Rauh, Olbrich, Jost, & Ay, 2014; Makkeh, Theis, & Vicente, 2018; Williams & Beer, 2010) for the analysis of synergistic, redundant, and unique information two source nodes have about one target node. Where applicable, IDTxl provides the option to return local variants of estimated measures (Lizier, 2014a). Also, tools are included for group-level analysis of the inferred networks, e.g. comparing between subjects or conditions in neural recordings.

The toolkit is highly flexible, providing various information-theoretic estimators for the user to select from; these handle both discrete and continuous time-series data, and allow choices, e.g. using linear Gaussian estimators (i.e. Granger causality, Granger (1969)) for speed versus nonlinear estimators (e.g. Kraskov, Stögbauer, & Grassberger (2004)) for accuracy (see the IDTxl homepage for details). Further, estimator implementations for both CPU and GPU compute platforms are provided, which offer parallel computing engines for efficiency. IDTxl provides these low-level estimator choices for network analysis algorithms but also allows direct access to estimators for linear and nonlinear estimation of (conditional) mutual information, TE, and AIS for both discrete and continuous data. Furthermore low-level estimators for the estimation of PID from discrete data are provided.

The toolkit is a next-generation combination of the existing TRENTOOL (Lindner, Vicente, Priesemann, & Wibral, 2011) and JIDT (Lizier, 2014b) toolkits, extending TRENTOOL's pairwise transfer entropy analysis to a multivariate one, and adding a wider variety of estimator types. Further, IDTxl is Python3 based and requires no proprietary libraries. The primary application area for IDTxl lies in analysing brain imaging data (import tools for common neuroscience formats, e.g. FieldTrip, are included). However, the toolkit is generic for analysing multivariate time-series data from any discipline. This is realised by providing a generic data format and the possibility to easily extend the toolkit by adding import or export routines, by adding new core estimators, or by adding new algorithms based on existing estimators.